# An Improved Masking Strategy for Self-supervised Masked Reconstruction in Human Activity Recognition

Jinqiang Wang[1], Tao Zhu[1,*] and Huansheng Ning[2]

## Abstract

Masked reconstruction serves as a fundamental pretext task for self-supervised learning, enabling the model to enhance its feature extraction capabilities by reconstructing the masked segments from extensive unlabeled data. In human activity recognition, this pretext task employed a masking strategy centered on the time dimension. However, this masking strategy fails to fully exploit the inherent characteristics of wearable sensor data and overlooks the inter-channel information coupling, thereby limiting its potential as a powerful pretext task. To address these limitations, we propose a novel masking strategy called Channel Masking. It involves masking the sensor data along the channel dimension, thereby compelling the encoder to extract channel-related features while performing the masked reconstruction task. Moreover, Channel Masking can be seamlessly integrated with masking strategies along the time dimension, thereby motivating the self-supervised model to undertake the masked reconstruction task in both the time and channel dimensions. Integrated masking strategies are named Time-Channel Masking and Span-Channel Masking. Finally, we optimize the reconstruction loss function to incorporate the reconstruction loss in both the time and channel dimensions. We evaluate proposed masking strategies on three public datasets, and experimental results show that the proposed strategies outperform prior strategies in both self-supervised and semi-supervised scenarios.

## 1. Introduction

The proliferation of wearable devices has facilitated easy access to human activity data. This accessibility has sparked significant interest in human activity recognition (HAR) based on wearable devices, propelling the rapid advancement of pervasive computing [1]. This technology finds human-centered applications in diverse domains such as healthcare [2-4], sports assessment [5, 6], and smart homes [7, 8]. Deep learning methods [9-12] have emerged as a pivotal driver in enhancing the performance of human activity recognition when compared to traditional machine learning approaches [13-15]. These deep learning methods leverage deep neural networks to process sensor data and infer activity categories through classifiers. However, most deep learning methods follow the supervised learning paradigm, which requires abundant labeled data to train a model. Unfortunately, manually labeling sensor data proves to be an arduous and time-consuming task, presenting a formidable challenge that restricts the progress of human activity recognition [16].

*Corresponding Author: Tao Zhu (tzhu@usc.edu.cn)
[1] School of Computer Science, University of South China, Hengyang, China
[2] School of Computer and Communication Engineering, University of Science and Technology Beijing, Beijing China
Codes are available at https://github.com/diheal/channle_masking

To address the challenge of the scarcity of labeled data in HAR, a promising solution has emerged: self-supervised learning [17]. This solution alleviates the issue of limited labeled data by formulating and addressing self-generated tasks on expansive datasets that do not necessitate human annotations. Capitalizing on the inherent structure and information within the data, self-supervised learning guides the learning process, empowering the encoder to capture valuable feature representations. Currently, several self-supervised learning pretext tasks have been successfully applied to HAR, each yielding encouraging results. These tasks include augmentation method discrimination [18], contrastive learning [19-22], data reconstruction [23], and masked reconstruction [24, 25].

Existing self-supervised models with masked reconstruction are randomly masking sensor data along the time dimension [24, 25]. As illustrated in Fig.1(a), the sensor data is subject to random masking along the time axis. Additionally, Fig. 1(b) demonstrates the application of Span Masking, where contiguous segments of sensor data along the time axis are randomly masked. After masking the data, the self-supervised model endeavors to reconstruct the sensor data, incorporating the masked positions. Upon the completion of pretraining, the encoder of the self-supervised model is deployed for the downstream activity classification task.

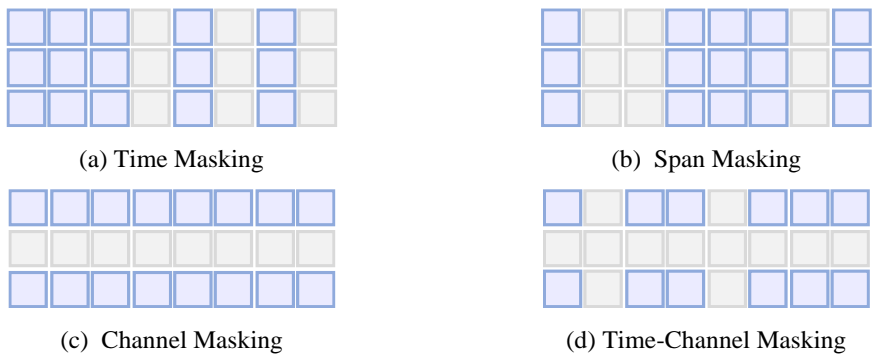

**Fig. 1.** Diagram of different masking strategies. The above example is a segment of sensor data with time on the horizontal axis and channels on the vertical axis. The gray block indicates the masked portion.

However, the masking strategy along the time dimension does not comprehensively account for the characteristics of the sensor data and ignores the importance of the information interactions between the channels. Consequently, the encoder's feature extraction capability may be hindered, particularly when facing sensor channel anomalies.

To address this limitation, we propose a novel masking strategy called Channel Masking. As depicted in Fig.1(c), sensor data is randomly masked along the channel dimension. This strategy can motivate self-supervised learning models to extract channel-dependent information. Furthermore, Channel Masking can be combined with masking strategies along the time dimension. The combined masking strategies are named Time-Channel Masking and Span-Channel Masking. Such masking strategies would empower the self-supervised model with the ability to model both time and channel dimensions and enhance the encoder's ability to extract semantic features. Finally, owing to the change in masking strategy, we optimized the previous loss function so that it can reconstruct the loss in both time and channel dimensions, prompting the model to learn extracting both time-dependent and channel-dependent features.

We evaluate our masking strategies on three public datasets, namely USC-HAD, UCI-HAR, and MotionSense. Additionally, we assess the performance of these strategies under two experimental scenarios: self-supervised learning, where the downstream task uses a large amount of labeled data, and semi-supervised learning, where the downstream task uses a limited amount of labeled data. Experimental results show that our proposed strategies outperform previous masking strategies. Furthermore, this paper explores the sensitivity of the hyperparameters of proposed masking strategies and the effect of whether the same mask position is used for the same training batch. Finally, this paper evaluates Channel Masking in channel anomaly experiments, and its experimental results show that the self-supervised model with Channel Masking can significantly mitigate the channel anomaly problem. This paper contributes the following.

1. This paper considers firstly the importance of sensor channel-dependent features in the masked reconstruction task and proposes a new masking strategy called Channel Masking. It can be merged with masking strategies along the time dimension called Time-Channel Masking and Span-Channel Masking.

2. Our proposed masking strategies outperform previous masking strategies on three public datasets and two experimental scenarios.

3. We optimize the reconstruction loss function to be compatible with both channel dimension and time dimension reconstruction losses.

4. This paper demonstrates that our proposed masking strategy is effective in mitigating the channel anomaly problem compared to supervised learning.

The remainder of the paper is structured as follows. Section 2 summarizes the related work; Section 3 describes in detail Channel Mask, Time-Channel Masking, and Span-Channel Masking, along with the optimized reconstruction loss functions. Section 4 describes the experimental setup for evaluating our proposed method; Section 5 shows the main experimental results and discussion; Section 6 provides further analysis and discussion on the details of the masking strategy; Section 7 summarizes the paper and proposes new challenges based on the identified shortcomings.

## 2. Related Work

### 2.1 Human Activity Recognition

Human activity recognition refers to the automatic identification of human activities through a computer system, encompassing actions such as walking, running, ascending and descending stairs, lying down, and falling. This technology utilizes sensors to capture individuals' behavioral patterns, and the collected sensor data undergoes processing within the computing system to perform activity recognition [26]. Commonly used sensors include accelerometers, gyroscopes, and magnetometers [27]. Numerous machine learning methods [13-15] have been employed to implement human activity recognition techniques. Deep learning methods with powerful feature extraction are gradually becoming mainstream in the field of activity recognition as machine learning requires manual selection of features and expert knowledge. Deep learning methods that follow the supervised learning paradigm [9, 11, 28, 29, 30] significantly improve the accuracy of activity recognition compared to traditional machine learning methods. However, these supervised learning methods usually require a large amount of annotated data to train a model, and the process of annotating data is time-consuming, tedious, and error-prone. Therefore, insufficient annotated sensor data is a major challenge for human activity recognition [16].

### 2.2 Self-supervised Learning

Self-supervised learning, also known as pretraining, involves accomplishing a pretext task by leveraging supervised signals generated in a non-human manner on a vast amount of unlabeled data [17]. The use of pretrained encoders has proven to be beneficial in enhancing performance in downstream tasks, effectively mitigating the challenge of limited labeled data availability. Several works have explored the application of self-supervised learning in human activity recognition. The research [18] uses data transformation recognition as the pretext task to train the self-supervised model. In the study [23], the authors applied multiple types of autoencoders to the pretraining of activity data. The study [31] applied CPC [32] for the first time to the pretraining of activity data, which resulted in superior performance compared to earlier approaches. Furthermore, the study [19] introduced contrastive learning to human activity recognition, investigating the impact of discriminative pretraining on performance. The study [20] improves activity recognition performance by employing the Transformer encoder for contrastive learning based on human activity recognition. The study [21] proposes a sensor data augmentation method based on contrast learning for enhancing the performance of pretrained models. The study [22] improves pretraining performance by optimizing the negative selection method in contrastive learning.

Masked reconstruction is a prevalent pretext task in self-supervised learning, aimed at encouraging the model to capture contextual features from the data. This pretext task finds applications in various domains. For instance, in natural language processing, one of the pretext tasks of BERT [33] and GPT [34] is to predict masked words. Similarly, in computer vision, the pretext task of MAE [35] is to reconstruct masked image patches. Masked reconstruction tasks have been effectively utilized in training self-supervised learning models for human activity recognition. The study [24] introduced the idea of BERT, a pretraining model in the field of natural language processing, using masked reconstruction as a pretext task to train the Transformer encoder to improve the performance of activity recognition. The study [25] also draws on the concept of BERT, employing Span Masking [36] to generate masks and sharing of parameters within the encoder to make the model lighter.

## 3. Proposed Methods

To fully harness the potential of self-supervised learning models utilizing masked reconstruction and to comprehensively account for the intrinsic characteristics of sensor data, we introduce an innovative masking strategy for sensor data, called Channel Masking. The procedural steps for implementing this masking strategy are delineated as follows.

A continuous segment of sensor data can be expressed in the form shown in Equation (1), where N is the length of time, K denotes the number of channels, and $x_i^j$ denotes the data of the jth channel at the ith moment. The Channel Masking operation is shown in Equation (2), where C is a randomly taken set of channel indices, and $I_{[j \in C]}$ is an indicator function, which is equal to 1 when the expression in [·] is true and 0 otherwise The Channel Masking schematic is shown in Fig.1(c).

$$x_i^j \quad i = 1,2, \dots, N; j = 1,2, \dots, K; \tag{1}$$

$$\tilde{x}_i^j = \left(1 - I_{[j \in C]}\right) \cdot x_i^j \quad i = 1,2, \dots, N; j = 1,2, \dots, K; \tag{2}$$

Channel Masking can be combined with masking strategies along the time axis as shown in Equation (3), where T is the set of indices of the time dimension to be masked. The data is masked along both time and channel dimensions. If T is generated by randomization [24], the combined masking strategy is named Time-Channel Masking. If T is generated by Span Masking [25][36], the combined masking strategy is called Span-Channel Masking.

$$\tilde{x}_i^j = (1 - I_{[j \in C || i \in T]}) \cdot x_i^j \quad i = 1,2, \dots, N; j = 1,2, \dots, K; \tag{3}$$

The above proposed masking strategies will be applied as the training process shown in Fig. 2. This framework is based on the improvement of the Mask Reconstruction framework and we focus on optimizing the data masking strategy. Given the incorporation of channel-related information within the masking strategies, the conventional reconstruction loss function becomes inapplicable to this new framework. Consequently, it becomes imperative to revise the reconstruction loss function, accommodating the introduced by both time and channel-based masking strategies.

The original mask reconstruction loss function is shown in Equation (4). Herein, $X_{time\ mask}^{raw}$ denotes the mask data along the time dimension in the original data, while $X_{time\ mask}^{rec}$ denotes the mask data along the time dimension in the reconstructed data. The term "MSE" signifies the mean square error loss. The loss function corresponding to our proposed Channel Masking strategy is shown in Equation (5), where only the reconstruction loss of the masked channel is computed. To incorporate both the Time Masking and Channel Masking strategies into the loss function, we merge $loss_{time}$ and $loss_{channel}$ by weighting as outlined in Equation (6). The parameter $\alpha \in [0,1]$ is introduced to measure the relative significance assigned to the time and channel dimensions within the model. Unless otherwise specified, α is set to 0.5.

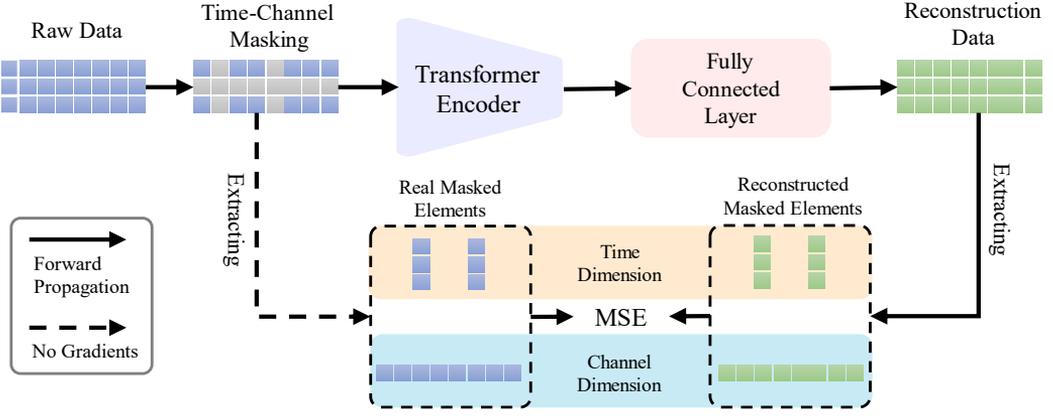

**Fig.2**. A self-supervised learning framework of Masked Reconstruction using Time-Channel Masking. The raw data is masked by masking strategies. The resulting masked data subsequently serves as input to the encoder, facilitating the extraction of intermediate features. These intermediary features are then directed into a fully connected layer, tasked with the reconstitution of the raw data. Following this stage, the difference in the masked positions between the reconstructed data and the raw data denoted as the reconstruction loss, is computed. Subsequently, during the downstream task, only the encoder with frozen parameters is used.

$$loss_{time} = MSE(X^{raw}_{time\ mask}, X^{rec}_{time\ mask}) \tag{4}$$

$$loss_{channel} = MSE(X^{raw}_{channel\ mask}, X^{rec}_{channel\ mask}) \tag{5}$$

$$loss = \alpha * loss_{time} + (1 - \alpha) * loss_{channel} \tag{6}$$

## 4. Experimental Setup

### 4.1 Datasets

The USC-HAD [37] activity recognition dataset comprises accelerometer and gyroscope data extracted from 14 participants who engaged in 12 daily activities. These activities encompass walking forward, walking left, walking right, walking upstairs, walking downstairs, running, jumping, sitting, standing, lying down, and riding up and down in an elevator. The accelerometer and gyroscope data were captured utilizing a MotionNode device positioned in the volunteer's right hip. Subsequently, this data was transmitted to a laptop computer via wired connections. All data were collected at a 100 Hz sampling rate.

The UCI-HAR [38] activity recognition dataset encompasses accelerometer and gyroscope data that were collected from 30 volunteers aged 19 to 48 years. These individuals performed six fundamental activities, namely standing, sitting, lying down, walking, ascending, and descending stairs. Specifically, these data were collected utilizing a smartphone (Samsung Galaxy S II) equipped with an embedded accelerometer and gyroscope. These sensors were affixed to the waist of each volunteer, enabling the capture of 3-axis linear acceleration and 3-axis angular velocity. This acquisition was conducted consistently at a sampling rate of 50 Hz.

The MotionSense [39] activity recognition dataset is a collection of accelerometer data and gyroscope data from 24 volunteers of different genders, ages, weights, and heights. These volunteers undertook six basic activities: going downstairs, going upstairs, walking, jogging, sitting, and standing. The specific collection process involved placing a smartphone (iPhone 6s) in the front pocket of the volunteers. Subsequently, the accelerometer and gyroscope data were systematically collected via the Core Motion framework on iOS devices using SensingKit. The sampling frequency was steadfastly maintained at 50 Hz throughout the entire data collection process.

## 4.2 Self-supervised Experimental Settings

In consonance with prior studies [22, 24], the data from subjects 11-12 in the USC-HAD dataset were designated as the validation set, subjects 13-14 were designated as the test set, and the rest of the subjects were designated as the training set. Building on the established practices of [22, 38], 20% of the subjects in the UCI-HAR and MotionSense were randomly selected as the test set, 20% of the subjects in the rest of the data were used as the validation set, and the remaining subjects' data were used as the training set. Based on the experience of [20, 24], in the USC-HAD and MotionSense datasets, the data were cut according to a one-second sliding window with a 50% overlap between the windows. Based on the experience of [22, 38], in the UCI HAR dataset, 128 readings were sampled as a sliding window with a 50% overlap between the sliding windows.

The structure of the self-supervised model is as follows:

Encoder: The initial component entails a singular layer of 1x1CNN, generating an output channel count of 128. Subsequently, this is succeeded by a Transformer-Encoder network which uses cosine positional coding, attention block is 3 layers and the number of heads of multi-head attention is 4.

Fully Connected Layer: It consists of three layers of MLP with the number of neurons accordingly by 256, 128, and the number of data channels. Between every two layers, there exists Batch Normalization, ReLU, and Dropout with a dropout rate of 0.1.

The pretraining procedure is to train the self-supervised model for 150 epochs. Each epoch employs a Batch Size of 256, utilizing the Adam optimizer with an initial learning rate of 1e-3. The loss function presented in Equation (4) is applied to the time dimension masking strategy, while the loss function elucidated in Equation (5) is employed for the channel dimension masking strategy. Furthermore, the loss function expounded in Equation (6) is used for the integrated masking strategy. Note that a training batch uses the same mask position generated randomly.

For the downstream task, the encoder acquired from the pretraining phase is employed with all layers frozen. Subsequently, randomly initialized fully connected layers are introduced with the number of neurons of the last MLP layer is the number of activity categories. The model is trained on a labeled training set and the performance of the pretrained model is evaluated on a test set. The mean F1-score [40] was used as the evaluation metric. The model was trained for 100 epochs with a Batch Size of 1024. Adam [41] optimizer with an initial learning rate of 1e-3 was used to optimize the model. Cross entropy loss is used as the loss function.

Since the test sets of the UCI-HAR and MotionSense datasets were obtained by random division, we sampled the datasets five times and trained them separately, and the final results were averaged. In this paper, each experimental result is obtained by conducting 5 individual experiments and then calculating the average value.

## 4.3 Semi-supervised Experimental Settings

In the semi-supervised experimental setup, the process and parameters of pretraining the model are the same as in the self-supervised experimental setup. In the downstream task, to reflect the shortage of labeled data, x labeled samples per class are sampled from the training set to be used as the training set, where $x \in \{1, 2, 5, 10, 25, 50, 100\}$. The validation set and test set settings are identical to the self-supervised setup. The rest of the experimental setup follows the self-supervised experimental setup unless otherwise stated.

# 5. Experimental Result

## 5.1 Self-supervised Experiment

In this subsection, we conduct an empirical evaluation to assess the efficacy of self-supervised models employing various masking strategies within the context of an activity recognition task. Specifically, we consider masking ratios of 10% for Time Masking and 15% for Span Masking based on the experience of the work [24, 25]. Our proposed Channel Masking introduces a 50% masking ratio, meaning that for USC-HAD and MotionSense datasets, 3 randomly chosen channels are masked, and for UCI-HAR, 5 channels are masked. For Time-Channel Masking, the ratio of time dimension masking is 10% and the number of channel masks is 2 on all datasets. For Span-Channel Masking, the ratio of time dimension masking is 15% and the number of channel masks is 2 on all datasets. Here, the purpose of combining masks is to explore whether the performance of the model is improved by adding channel masks based on time masks. The performance of the supervised learning model is also compared, which differs from the self-supervised learning model in that the encoder is randomly initialized instead of freezing the pretraining parameters.

**Table 1.** Performance of pretrained models using various masking strategies on a human activity recognition task. The evaluation metric is F1-score. "Supervised" denotes a supervised learning model. The rest of the experimental results follow the self-supervised experimental setup.

| Methods | USC-HAD | UCI-HAR | MotionSense |
| --- | --- | --- | --- |
| Supervised | 58.43 | 92.81 | 84.45 |
| Masking Strategies | | | |
| Time Masking [24] | 50.39 | 79.24 | 82.79 |
| Span Masking [25] | 49.32 | 89.23 | 83.17 |
| Channel Masking (ours) | 61.49 | 92.26 | **84.93** |
| Time-Channel Masking (ours) | **62.20** | 92.34 | 84.28 |
| Span-Channel Masking (ours) | 58.25 | **92.76** | 83.96 |

The experimental results presented in Table 1 show that our proposed masking strategies exhibit superior performance compared to the previous strategies across three distinct datasets. This observation demonstrates that prompting the self-supervised learning model to consider channel-related information can improve its performance on activity recognition tasks. On the three datasets, the difference between the performance of our proposed optimal masking strategy and the performance of the optimal previous masking strategy is 11.81, 3.53, and 1.76, respectively. These differences suggest that the performance of the self-supervised learning framework based on masked reconstruction is highly dependent on the choice of masking strategy. In the comparative analysis with supervised learning, the self-supervised learning model beats supervised learning on the USC-HAD dataset only and is slightly inferior to the supervised learning model on the UCI-HAR and MotionSense datasets. This divergence may arise from the exceptional performance of the supervised learning model on UCI-HAR and MotionSense datasets, constraining the potential enhancement of the self-supervised approach. Conversely, the USC-HAD dataset reveals suboptimal performance by the supervised learning model, affording ample room for improvement by the self-supervised paradigm.

### 5.2 Semi-supervised Experiment

Here, we explore the impact of different masking strategies for the self-supervised learning framework in the setting of limited labeled data. The experimental setup here follows the semi-supervised experimental setup of Section 4.3. The supervised learning benchmarks are also compared with the same model structure as presented in the downstream task of the self-supervised model, with the difference that its parameters are randomly initialized. The experimental results are shown in Fig.3.

Experimental results show that our proposed masking strategies consistently surpass previous masking strategies across all three datasets and seven labeling ratios. The associated confidence intervals conspicuously establish its statistically significant superiority. This observation suggests that the self-supervised learning model considering channel-related information can improve activity classification performance in a restricted labeling environment. Notably, the self-supervised model, employing our devised masking strategies, outperforms the supervised learning benchmark. Particularly, for the USC-HAD dataset, it can be seen by the confidence intervals that our proposed masking strategies significantly outperform supervised learning models, while the prior masking strategies significantly trail behind the supervised learning models. This once again proves the importance of our proposed strategies. On the UCI-HAR and MotionSense datasets, our proposed masking strategies exhibit marginal and statistically insignificant performance improvements over the supervised learning model across the majority of experimental setups. Conversely, our approaches are insignificantly worse than the supervised learning model for a small number of setups. This observation could be attributed to the intrinsic ease of training the UCI-HAR and MotionSense datasets, which potentially constrains the potential improvement space for the self-supervised learning model. Overall, in a limited labeled data setting, the self-supervised learning model exhibits more pronounced improvements relative to the supervised learning model, as opposed to situations with ample labeled data.

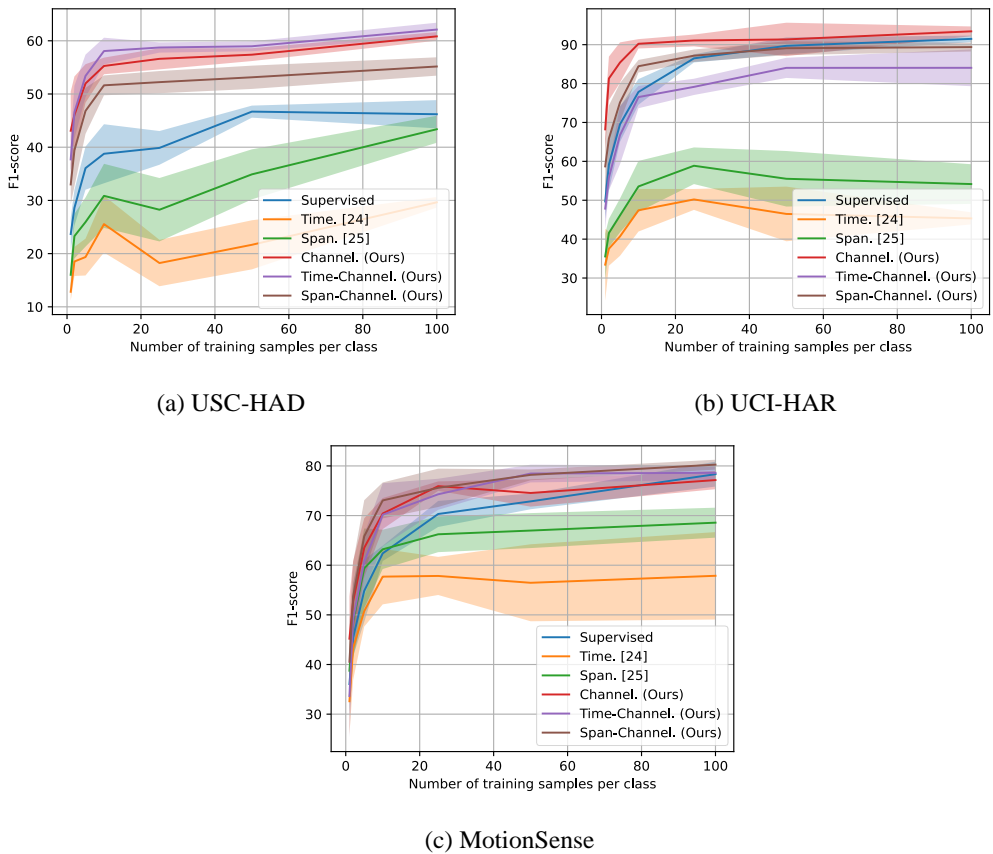

**Fig. 3.** The performance of self-supervised learning models using different masking strategies for activity classification in limited labeled data. The evaluation metric is F1-Score. All results are obtained by averaging five experimental results, and the shaded areas denote 95% confidence intervals.

## 6. Discussion

This section is dedicated to investigating the impact of the hyperparameters associated with the proposed masking strategies on model performance. A pretraining trick tailored for a self-supervised model based on masked reconstruction is also explored. According to previous experience [22], the smaller the number of tags, the higher the parameter sensitivity and the more intuitive the corresponding performance difference. Thus, the experimental setup in this section follows the semi-supervised experimental setup, with the only difference being that in the downstream task, the sampling size of each class randomly sampled from the training set is fixed to 10.

## 6.1 Masking Ratio

The masking ratio exerts a direct influence on the complexity of the reconstruction task. The larger the masking ratio is, the less valid information is contained in the masked data, and the more difficult the reconstruction task is. For this reason, the effect of the masking ratio on the performance of the self-supervised model is explored. Here, the time and channel dimensions are explored separately, and the experimental results are shown in Tables 2 and 3.

**Table 2.** Classification performance of self-supervised learning models using different time masking ratios. The F1-score serves as the evaluation metric, with the designated masking strategy being Time Masking. The "Masking Ratios" denotes the proportion of masked sensor data along the time axis.

| Masking Ratios | USC-HAD | UCI-HAR | MotionSense |
| --- | --- | --- | --- |
| 5% | 7.15 | 41.95 | 52.22 |
| 10% | 22.17 | 32.48 | 49.46 |
| 20% | 28.35 | 50.12 | 58.40 |
| 30% | 17.74 | 51.11 | 57.80 |
| 50% | 19.71 | 56.69 | 65.67 |
| 70% | 24.86 | 61.53 | 64.79 |
| 90% | **39.31** | **65.59** | **70.07** |

**Table 3.** Classification performance of self-supervised learning models employing varied channel masking numbers. The evaluation metric employed is the F1-score, with the selected masking strategy being Channel Masking. The "Masking Numbers" signifies the number of masking sensor data channels. The number of data channels is 6 for the USC-HAD and MotionSense datasets and 9 for the UCI-HAR dataset.

| Masking Numbers | USC-HAD | UCI-HAR | MotionSense |
| --- | --- | --- | --- |
| 1 | 43.74 | 60.85 | 70.27 |
| 2 | 45.66 | 70.37 | 69.61 |
| 3 | **57.21** | 71.90 | **71.79** |
| 4 | 51.23 | **78.81** | 68.70 |
| 5 | 37.24 | 78.53 | 60.44 |

Time Masking is discussed first. As the degree of masking amplifies, there is a concurrent enhancement observed in the overall model performance across all three datasets. Notably, there are only transient and localized performance dips. On all three datasets, the most effective ratio of masking is determined to be 90%. This revelation diverges from established conventions rooted in earlier research, where prevalent practice involved the utilization of default masking ratios of either 10% or 15%. This empirical finding demonstrates that a self-supervised learning framework based on mask reconstruction can adapt to large-scale time masks and outperforms small-scale time masks.

Channel Masking is discussed second. On both the USC-HAD and MotionSense datasets, the model performance is increasing and then decreases as the masking number increases. The best model performance corresponds to a masking number of 3, which is exactly half the number of sensor data channels. On the UCI-HAR dataset, optimal performance aligns with a masking number of 4, akin to the performance at a masking number of 5. Notably, the UCI-HAR dataset has a data channel number of 9, and the optimal masking number is also right around half of the number of data channels. The above discussion leads to the conclusion that leveraging Channel Masking with cardinality equal to half the data channels remarkably enhances self-supervised learning models rooted in masked reconstruction. This conclusion doubles as a practical guideline for training other sensor datasets.

Finally, by comparing Tables 2 and 3, it is found that the peak performance of the model with Channel Masking outperforms the best performance of the model with Time Masking. The findings again prove that the self-supervised learning model that considers channel-related information outperforms the self-supervised learning model that considers time-related information.

### 6.2 The hyperparameter α

Here, we delve into the significance of time reconstruction loss and channel reconstruction loss within the Time-Channel Masking approach. Specifically, we analyze the impact of parameter α in Equation 6 on model performance. To better understand the importance of these reconstruction losses, we maintain similar mask ratios for both time and channel aspects. For the USC-HAD and MotionSense datasets, the channel masking number was 1 (16.7%) and the time masking ratio was 17%. Likewise, for the UCI-HAR dataset, the channel masking number is 2 (22.2%) and the time masking ratio is 22%. Detailed experimental results are presented in Table 4.

**Table 4.** Effect of varying hyperparameters α on model performance using the Time-Channel Masking strategy. Decreasing α corresponds to a lower percentage of time reconstruction loss. The evaluation metric is F1-score.

| α | USC-HAD | UCI-HAR | MotionSense |
| --- | --- | --- | --- |
| 0.1 | 45.50 | 61.13 | 70.52 |
| 0.3 | **48.45** | 70.60 | **70.79** |
| 0.5 | 47.71 | **77.70** | 69.42 |
| 0.7 | 44.41 | 71.44 | 44.41 |
| 0.9 | 46.57 | 68.25 | 46.57 |

The experimental results show that in the USC-HAD and MotionSense datasets, the best performance corresponds to an α of 0.3, meaning that the model performs best when the channel reconstruction loss accounts for a large portion of the model. This empirical validation further emphasizes the importance of channel-dependency information for self-supervised learning models based on masked reconstruction. On the UCI-HAR dataset, the best performance corresponds to an α of 0.5, namely, the model performs best when the channel reconstruction loss and the time reconstruction loss account for the same percentage. With the above analysis, the value of the hyperparameter α can be assigned to be slightly less than 0.5 when pretraining a new dataset. In other words, the weight of the channel reconstruction loss is set to be slightly higher than the time reconstruction loss.

### 6.3 Sensor Channel Anomaly Issues

Here, we explore whether the self-supervised learning model after pretraining using the Channel Masking strategy has the ability to handle data channel anomalies. To keep the generality intact, the masking number for Channel Masking is set to 3 in the pretraining. In this task, pretrained models are trained with normal data and evaluated using channel anomaly data. Furthermore, a comparative analysis is conducted against supervised learning benchmarks, the architectural description of which is presented in Section 5. The experimental results are presented in Table 5.

Experimental results show that the self-supervised learning model with Channel Masking surpasses the supervised learning model across the UCI-HAR and MotionSense datasets. This demonstrates that the model employing masked reconstruction of channel dimensions as a pretext task can improve the processing ability for channel anomaly data. On the USC-HAD dataset, the self-supervised learning model outperforms the supervised learning model when the masking number is 1 and 3, while inferior to the supervised model when the number of masks is 5. The divergence could potentially be attributed to the number of masked channels used in the pretraining is 3, which caused the model to overfit to this masking number. It is worth noting that pretraining was done using a masking channel number of 3, but when evaluated on anomalous data with a masking number of 1, it outperforms supervised learning on all three datasets, which suggests a certain degree of generalizability of the Channel Masking strategy.

**Table 5.** Evaluating the Efficacy of the self-supervised learning model based on mask reconstruction for mitigating data channel anomalies. "Masking Number" indicates the number of abnormal channels in the test set, where the abnormal value consistently set to 0. "Supervised" denotes the supervised learning model and "Self-supervised" denotes the self-supervised learning model with Channel Masking. The F1-score serves as the designated evaluation metric.

| Masking Number | Paradigms | USC-HAD | UCI-HAR | MotionSense |
| --- | --- | --- | --- | --- |
| 1 | Supervised | 45.87 | 73.87 | 51.25 |
| | Self-supervised | **60.40** | **79.53** | **75.07** |
| 3 | Supervised | 37.63 | 68.24 | 44.32 |
| | Self-supervised | **47.83** | **77.67** | **62.91** |
| 5 | Supervised | **22.04** | 55.37 | 32.46 |
| | Self-supervised | 18.31 | **65.86** | **34.59** |

### 6.4 A Pretraining Trick

Here, we explore a trick in pretraining a self-supervised learning model based on masked reconstruction that uses the same mask position for the same batch of data. Whereas the previous method has different mask positions between each sample within the same training batch. Now the pretraining uses two tricks, same position masking and different position masking, for the same batch of samples, with Time-Channel Masking as the masking strategy. And then evaluates the performance of the two tricks in the downstream activity classification task separately. Detailed results can be found in Table 6.

**Table 6.** The effect of the same batch of data using the same masking and different masking tricks on model performance. Time-Channel Masking was set on the USC-HAD dataset with a time masking ratio of 20 and a channel masking number of 1. On the UCI-HAR and MotionSense datasets, the time masking ratio was set to 10 and the number of channels masked was 2. The above masking ratios were not intentionally set.

| Tricks | USC-HAD | UCI-HAR | MotionSense |
| --- | --- | --- | --- |
| Different. | 36.37 | 62.33 | 62.38 |
| Same. (ours) | **49.99** | **74.98** | **72.03** |

Empirical findings reveal that across the three datasets, the performance of the model using the same mask position for the same batch is higher than that of the model using different mask positions. The differences between the two tricks on the three datasets are 13.62, 12.65, and 9.65, respectively, demonstrating the trick of the same mask position can substantially improve the performance of the pretrained mode. This trick can also be utilized in subsequent self-supervised learning methods that conform to masked reconstruction.

## 7. Conclusion

To maximize the potential of self-supervised learning models based on mask reconstruction for human activity recognition tasks, this paper proposes a new masking strategy Channel Masking, coupled with a time dimension masking strategy, yielding Time-Channel Masking and Span-Channel Masking. Proposed strategies are rigorously evaluated on three public datasets, encompassing both self-supervised and semi-supervised experimental scenarios. Results substantiate the superiority of the proposed masking strategies over preceding strategies, underscoring the pivotal role of channel-dependent information. Furthermore, the proposed Channel Masking strategy is evaluated in sensor channel anomaly experiments, affirming Channel Masking can enhance the model's capability to handle data channel anomalies. Finally, the hyperparameters of the proposed masking strategies and a training trick are explored, and relevant recommendations are given based on the experimental results.

In addition, our study does exhibit certain limitations. When confronted with a substantial volume of annotated data, the self-supervised model employing proposed masking strategies does evince certain disparities when juxtaposed with its supervised counterpart. This gap is also a problem shared by other self-supervised learning models. We will bridge this gap by improving the reconstruction effect and eliciting other supervised signals in future work.

### Author's Contributions

Conceptualization, TZ. Investigation and methodology, JW. Formal analysis and Supervision TZ, HN. Writing of the original draft, JW. Writing of the review and editing, TZ, HN. Funding acquisition TZ, HN.

### Funding

This work was supported in part by the National Natural Science Foundation of China (62006110, 62071213), the Natural Science Foundation of Hunan Province (2021JJ30574), the Research Foundation of Education Bureau of Hunan Province (21C0311, 21B0424), and Hengyang Science and Technology Major Project (202250015428).

### Competing Interests

The authors declare that they have no competing interests.